\documentclass[twocolumn,times,trackchanges, dvipsnames]{aastex631}
\usepackage[utf8]{inputenc}
\usepackage{amsmath}

\begin{document}
\title{The Hot Inner AU of V883 Ori}

\author[0000-0002-9540-853X]{Adolfo S. Carvalho}
\affiliation{Department of Astronomy; California Institute of Technology; Pasadena, CA 91125, USA}

\author{Lynne A. Hillenbrand}
\affiliation{Department of Astronomy; California Institute of Technology; Pasadena, CA 91125, USA}
\author[0000-0001-7157-6275]{Ágnes Kóspál}
\affiliation{Konkoly Observatory, HUN-REN Research Centre for Astronomy and Earth Sciences, MTA Centre of Excellence, Konkoly-Thege Mikl\'os \'ut 15-17, 1121 Budapest, Hungary}
\affiliation{Max Planck Institute for Astronomy, Königstuhl 17, D-69117 Heidelberg, Germany}
\affiliation{Institute of Physics and Astronomy, ELTE E\"otv\"os Lor\'and University, P\'azm\'any P\'eter s\'et\'any 1/A, 1117 Budapest, Hungary}
%\affiliation{\lah{do we need *all* of these affiliations as opposed to just the one (or two) where the work in this particular paper was actually carried out?}}

\begin{abstract}
    The V883 Ori system is a rapidly accreting young stellar object that has been used as a laboratory for studying the molecular inventory of young circumstellar disks with high luminosity. We simultaneously fit high resolution spectroscopy and medium resolution spectrophotometry of the system to constrain the physical conditions in the inner au. Using our thin viscous accretion disk model, we find $\dot{M} = 10^{-3.9 \pm 0.2} \ M_\odot$ yr$^{-1}$, $R_\mathrm{inner} = 5.86 \pm 1 \ R_\odot$, $i = 38.2 \pm 3$ degrees and $A_V = 20.8 \pm 0.7$ mag, resulting in an accretion luminosity of 458 $L_\odot$ and maximum disk temperature of 7045 K. The optical portion of the SED greatly exceeds the flux level expected for a highly extincted accretion disk. 
    We propose that the excess emission arises from a contribution due to scattering of the accretion disk spectrum off nearby envelope material that is viewed along a less-extincted line of sight. Additionally, we use photometric observations spanning 137 years to demonstrate that the source has accreted at least 18 $M_\mathrm{Jup}$ of disk material to date. Finally, we discuss the importance of considering both the viscous heating from the midplane and the consequent irradiation effects on the outer disk when modeling the temperature structure to reproduce millimeter-wavelength observations.
\end{abstract}

\section{Introduction}\label{sec:introduction}
There is a population of accreting young stellar objects (YSOs) known to undergo extreme photometric outbursts, during which the mass accretion rate onto the star increases by up to 10,000, causing the system to brighten by a factor of 100. The class prototype was defined when FU Ori itself rapidly brightened in 1937 by 6 magnitudes \citep[in $B$ band,][]{Wachmann_FUOri_1954ZA} and then remained at its elevated brightness for several decades \citep{herbig_eruptive_1977}. By the 1990s, around 10 FU Ori-like objects (FUOrs) had been discovered and their spectra were shown to be consistent with those of thin, viscously-heated accretion disks \citep{Kenyon_FUOri_disks_1988ApJ, hartmann_fu_1996}. 

Among these first half dozen or so discovered FUOrs was the V883 Ori system. Although it was only identified in 1993 as a FUOr in outburst, it is one of the brightest and longest-lived accretion outbursts, having potentially peaked in brightness sometime before 1888 \citep{Pickering_V883OriNebula_1890AnHar,StromStrom_V883OriDiscovery_1993ApJ}.

V883 Ori has been the target of several millimeter and submillimeter observations aimed at studying the molecular inventory of the disk. Due to the intense heat of the outburst, complex molecules that are typically trapped in ices in the midplane of the disk are released when the ices sublimate. The liberated molecules, now in gas phase, emit brightly at millimeter/submillimeter wavelengths. Although the details of this sublimation, like the new location of the H$_2$O snow line \citep{Schoonenberg_V883Ori_Not40auSnowline_2017A&A}, are not well known, the rich millimeter emission line spectrum of V883 Ori is well documented \citep{Cieza_ALMA_V883Ori_2016Natur,VantHoff_V883Ori_2018ApJ,Lee_V883Ori_COMs_2019NatAs, Tobin_v883Ori_2023Natur}.

Modeling the emission from these molecules to interpret the millimeter spectra requires a detailed understanding of the radiation field in the disk. In non-outbursting YSOs, the majority of the relevant radiation is from the central star (in the visible/near-infrared) and the accretion columns shocking on the stellar surface \citep[in the X-ray/ultraviolet,][]{oberg_ppvii_2023ARA&A}. In FUOrs, however, the star itself contributes minimally to the overall spectrum, while the magnetospheric accretion columns are expected to have been overwhelmed by the disk and thus also do not contribute much \citep{hartmann_fu_1996, Liu_fuorParameterSpace_2022ApJ}. 

Instead, the near-ultraviolet (NUV), visible, and near-infrared (NIR) spectrum is dominated by emission from the viscously heated accretion disk itself. The far-UV \citep{Carvalho_FUVSpectrumFUOri_2024ApJ} and X-ray \citep{kuhn_comparison_2019} are likely dominated by emission near the star-disk interface, either from enhanced magnetic activity during the outburst or a shock from the accretion flow along the surface of the disk.
In V883 Ori, most of the NUV/visible flux is faint due to the high line-of-sight extinction to the central source. This has limited the ability to constrain the properties of the inner disk like the disk mass accretion rate, accretion luminosity, and maximum temperature. 

\begin{figure}[htb]
    \centering
    \includegraphics[width=0.99\linewidth]{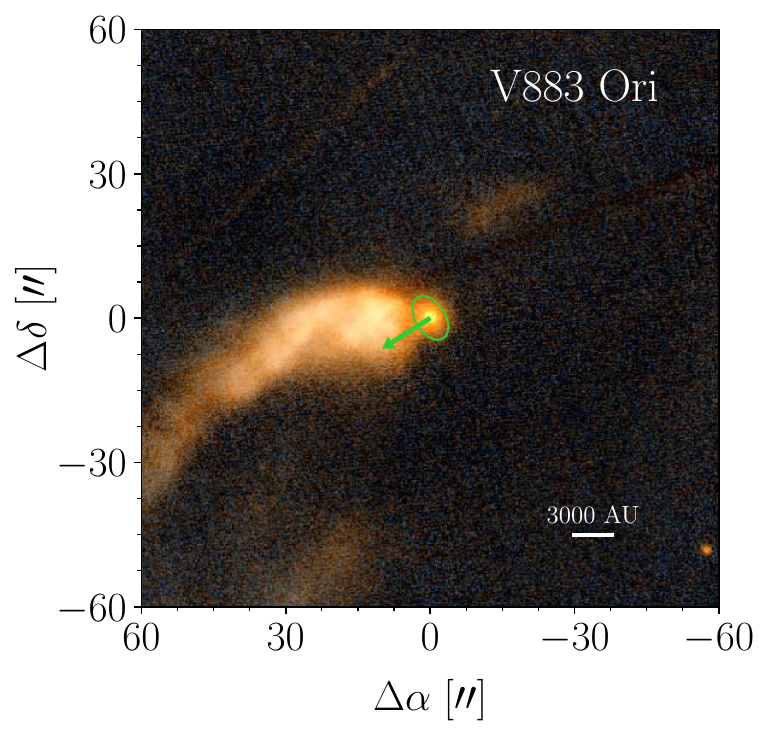}
    \caption{V883 Ori as seen in the Pan-STARRS $g-z$ color image. The green ellipse shows the projected shape of the disk on the sky (scaled $\times 12$ for visibility), assuming an inclination of 38.3$^\circ$ and a position angle of 32.4$^\circ$ \citep{cieza_v883Ori_2018MNRAS}. The green arrow shows the axis of rotation of the disk \citep{Lee_V883Ori_COMs_2019NatAs}. The position angle and inclination of the disk indicate it is the primary illumination source for the bright nebula that extends $\sim 70^{\prime\prime}$ to the southeast. The nebula is known as IC 430.}
    \label{fig:V883OriEnv}
\end{figure}

The V883 Ori system was first identified as a potential YSO by \citet{Haro_HalphaCatalog_1953ApJ}, who observed the morphological similarity of the bright nebula that extends $67^{\prime\prime}$ to the southeast of the point source (see Figure \ref{fig:V883OriEnv}) to Herbig-Haro objects. The point source, which is V883 Ori, was then classified as a B-type star under 20 mag of extinction by \citet{Allen_V883Ori_1975MNRAS}. Eventually, \citet{StromStrom_V883OriDiscovery_1993ApJ} highlighted the spectroscopic similarities between V883 Ori and FU Ori in low resolution optical spectra, identifying it as a FUOr. The FUOr classification of the source has since been re-affirmed by \citet{connelley_near-infrared_2018}. 

Millimeter continuum observations of V883 Ori have revealed a large outer disk, extending to 100 au, with an inclination of 38.3$^\circ$ and a position angle on the sky of 32.4$^\circ$ \citep{Cieza_ALMA_V883Ori_2016Natur}. The orientation of the disk relative to the bright nebula is shown in Figure \ref{fig:V883OriEnv}. 

In this article, we present new measurements of the inner disk temperature, luminosity, and truncation radius, enabled by deep spectroscopic observations of the source from $0.4-4.2$ $\mu$m. We demonstrate that the visible range emission from V883 Ori, while faint, is much brighter than expected for a highly extincted FUOr and likely due to scattering in the surrounding nebula. We then discuss our model of the system and its implications for interpretation of the emission at longer wavelengths.

\section{Data} \label{sec:data}
To fully characterize the inner disk of V883 Ori, we combine several public and newly collected datasets spanning the 0.4 to 5 $\mu$m and wavelength range, including both and high and low resolution spectra. The data are described below. 
\subsection{High Resolution Spectroscopy}
\subsubsection{Visible (Keck/HIRES)}
Two high dispersion red-optical spectra of V883 Ori were obtained on 5 October 2003 and 10 December 2011 using the HIgh RESolution spectrograph \citep[HIRES,][]{Vogt1994} on the W.M. Keck Observatory's Keck I telescope. The 2003 spectrum was obtained from the Keck Observatory Archive (PI: R. White) and was processed by their automated pipeline, which uses the MAKEE pipeline reduction package written by Tom Barlow\footnote{ {\url{https://sites.astro.caltech.edu/~tb/makee/}}}. The 2011 spectrum was also obtained from the Keck Observatory Archive (PI: L. Hillenbrand), but was extracted and reduced using the HIRES data reduction pipeline in PypeIt \citep{prochaska_pypeit_2020JOSS}. 

Both spectra were continuum normalized using the asymmetric least-squares detrending code described in \citet{carvalho_V960MonSpectra_2023ApJ}. The spectra both have a resolution of $R \equiv \lambda/\Delta\lambda = 37,000$ and have reasonable signal-to-noise from 6400 \AA\ to 9000 \AA. V883 Ori is very faint in the visible ($V \sim 20$ mag, see Section \ref{sec:LC}), requiring prohibitively long exposure times for greater sensitivity at shorter wavelengths. The position angle (PA) of the slit in the 2003 (2011) spectrum was to 72$^\circ$ ($50^\circ$), so the nebula shown in Figure \ref{fig:V883OriEnv} is mostly excluded from the slit. Visual inspection of the 2D images confirm that there no extended nebular emission captured during the observations. 

\begin{figure}
    \centering
    \includegraphics[width=0.99\linewidth]{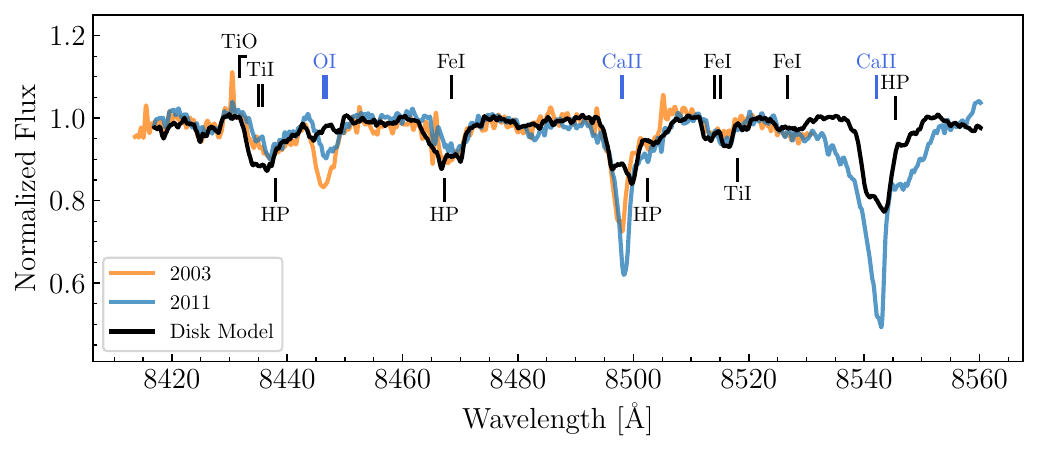}
    \includegraphics[width=0.99\linewidth]{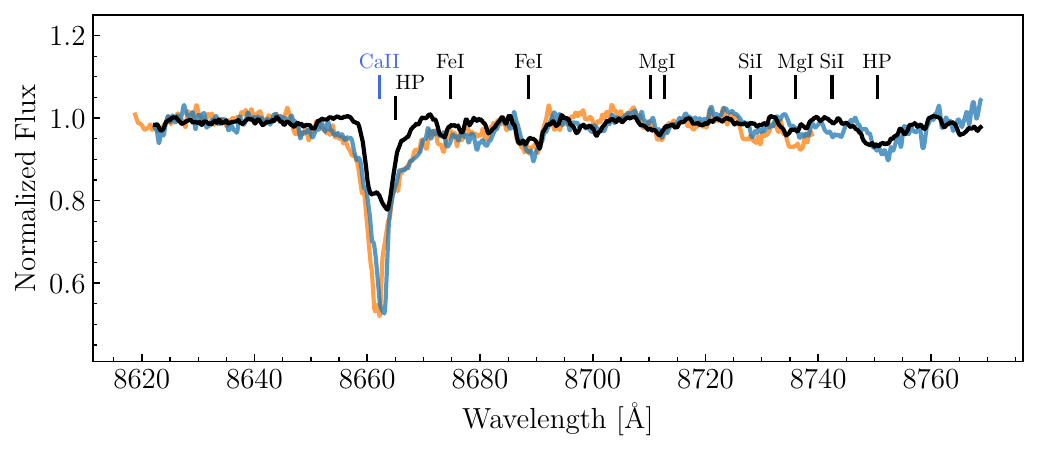}
    \includegraphics[width=0.99\linewidth]{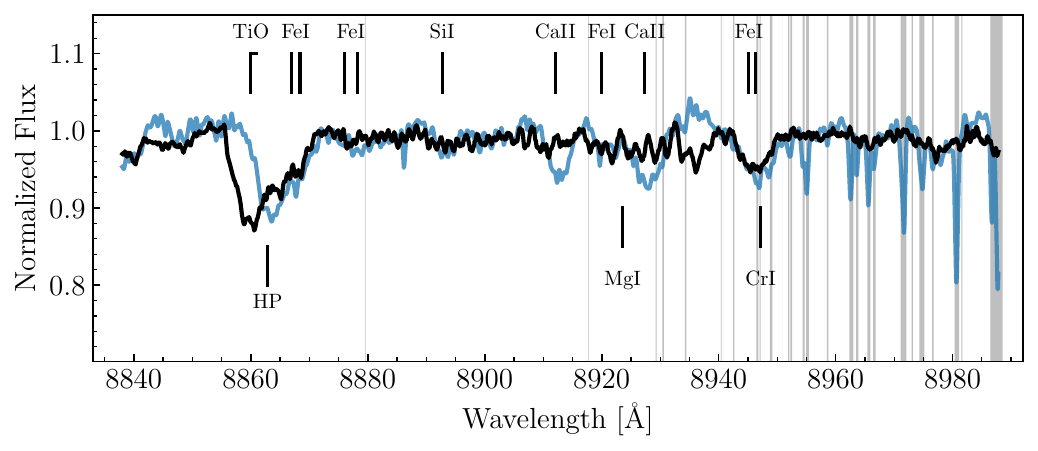}
    \caption{The reddest several orders of the 2003 (orange) and 2011 (blue) Keck/HIRES spectra, where the signal-to-noise is suitable, plotted along with the disk model described in Section \ref{sec:diskFits} (black). The model reproduces the weak absorption features well, but misses portions of several \ion{H}{1} and \ion{Ca}{2} profiles that arise in winds. Strong telluric features are present at the reddest wavelengths shown and are marked by grey vertical lines.}
    \label{fig:HIRES}
\end{figure}

Selected orders from the spectra are shown in Figure \ref{fig:HIRES}. Absorption features that trace the disk and those which trace the outflow in V883 Ori are marked in black and blue, respectively. The \ion{Ca}{2} IRT and \ion{O}{1} have blue-shifted absorption that is variable between 2003 and 2011, indicating they are outflow features tracing a wind in the inner disk \citep{Calvet_FUOriModel_1993}.

\subsubsection{NIR (Keck/NIRSPEC)}
We obtained a high resolution NIR spectrum of V883 Ori from $1-2.5 \ \mu$m on 31 Oct 2023 using the Keck Observatory's Near InfraRed SPECtrograph \citep[NIRSPEC,][]{McLean_nirspecDesign_1998SPIE, Martin_NIRSPECupgrade_2018SPIE10702E} using the NIRSPEC-1, NIRSPEC-3, NIRSPEC-5, and Kband-new filters to span the entire wavelength range. The exposure times for each band were 1200s, 480s, 360s, and 40s, respectively, to achieve a minimal signal-to-noise ratio (SNR) per pixel of 50 in each band. The slit PA was set to 0$^\circ$ to avoid the large nebula to the southeast. 

\begin{figure}
    \centering
    \includegraphics[width=0.99\linewidth]{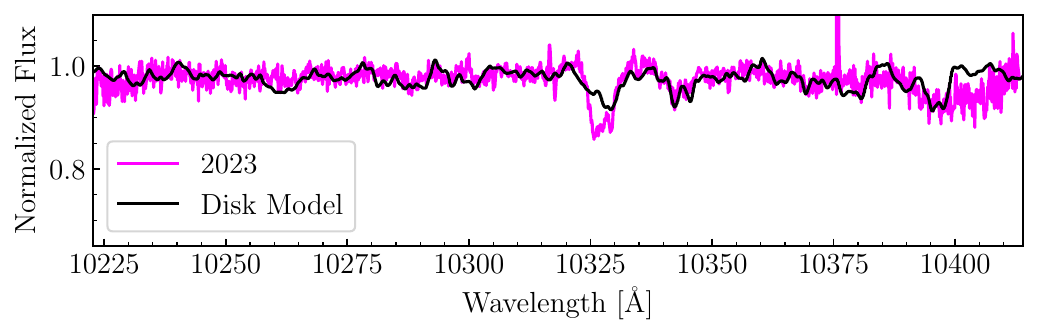}
    \includegraphics[width=0.99\linewidth]{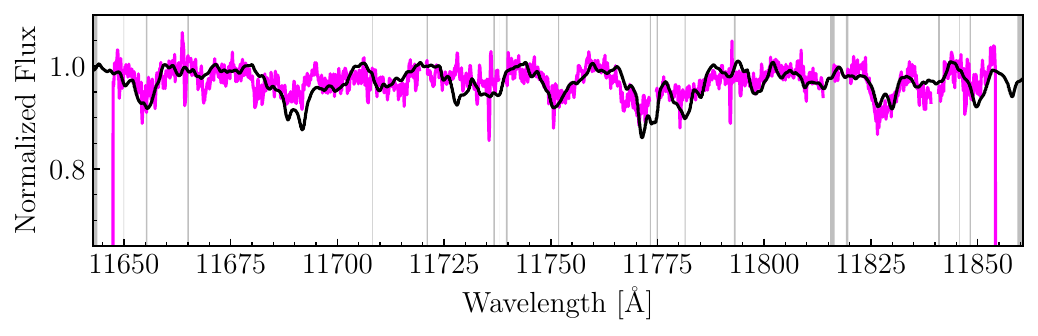}
    \includegraphics[width=0.99\linewidth]{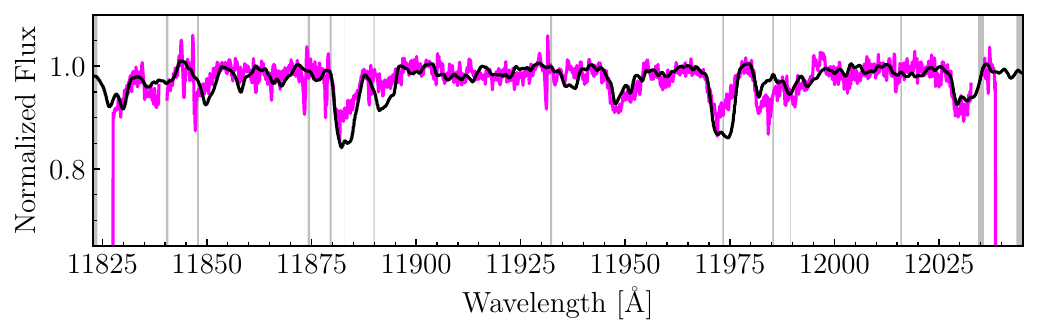}
    \includegraphics[width=0.99\linewidth]{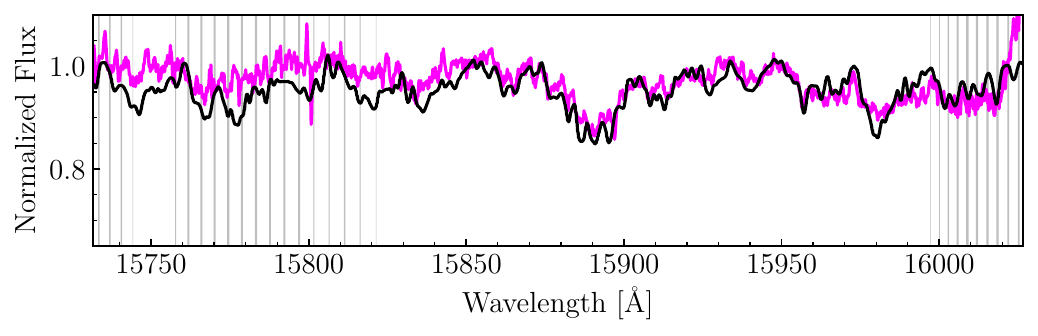}
    \includegraphics[width=0.99\linewidth]{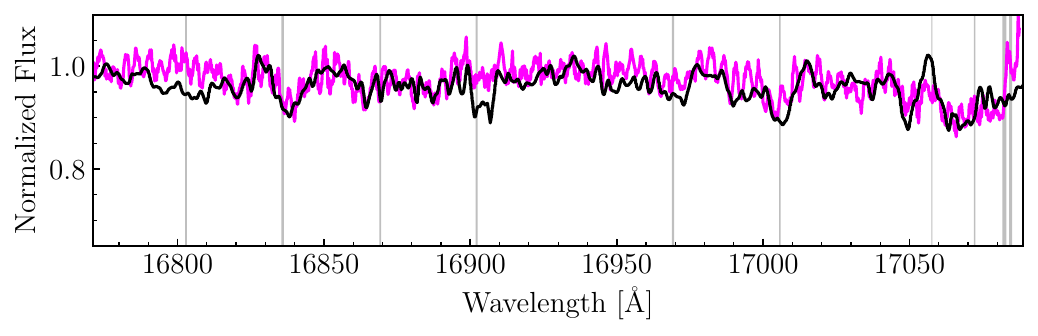}
    \includegraphics[width=0.99\linewidth]{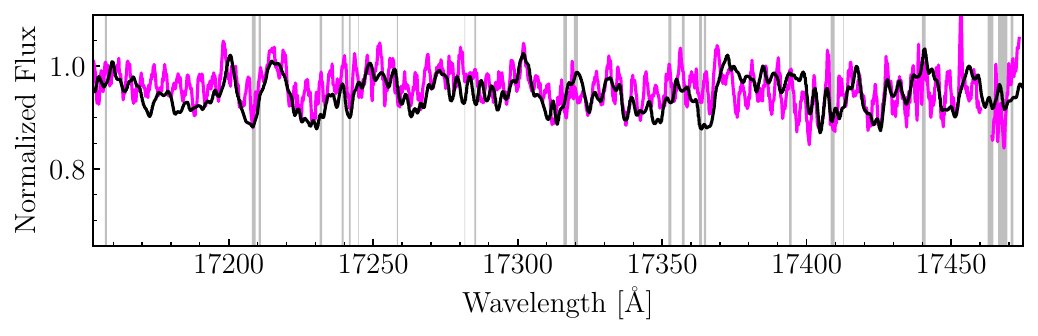}
    \includegraphics[width=0.99\linewidth]{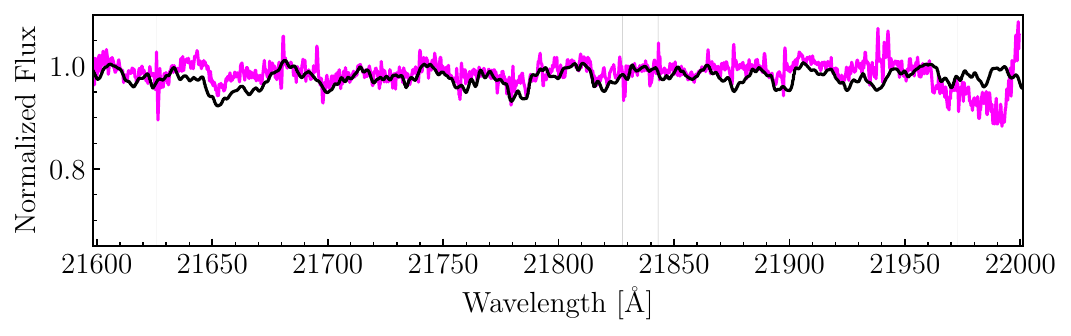}
    \includegraphics[width=0.99\linewidth]{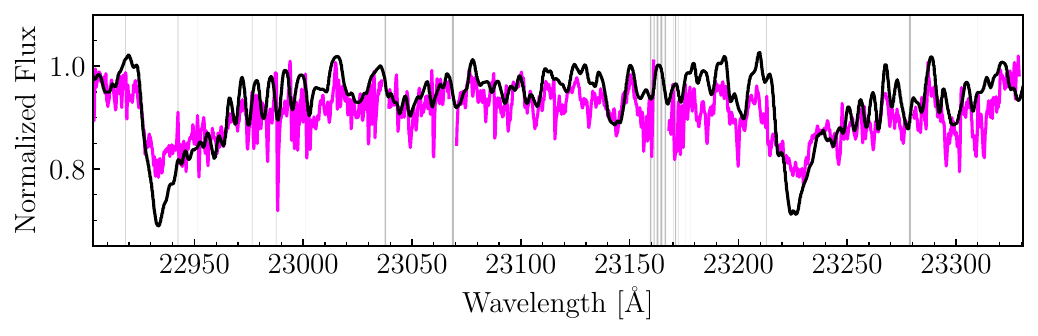}
    \caption{Select orders from the 2023 Keck/NIRSPEC spectrum (magenta) compared with the disk model described in Section \ref{sec:diskFits} (black). The model matches the majority of the absorption features accurately from 1.0 $\mu$ to 2.3 $\mu$m. Grey vertical bands mark regions of strong telluric absorption.}
    \label{fig:NIRSPEC}
\end{figure}

The spectra were extracted and reduced using the NIRSPEC pipeline \citep{Carvalho_PypeIt_2024RNAAS} in the data reduction software package PypeIt \citep{prochaska_pypeit_2020JOSS}. The telluric correction was performed using a high SNR spectrum of the A0V standard star HIP 102074, by first fitting an A0V template spectrum to remove the stellar hydrogen absorption, then continuum normalizing the remaining telluric spectrum to 1.0 (again using the asymmetric least-squares detrending code). We then divide the observed V883 Ori spectra by the normalized telluric spectrum. A description of the general procedure is given in \citet{Carvalho_PypeIt_2024RNAAS}. Select orders from the spectrum are shown in Figure \ref{fig:NIRSPEC}

\subsection{Spectrophotometry}
\subsubsection{NUV/Visible (Keck/LRIS)}
We obtained a visible range flux-calibrated spectrum of V883 Ori on 3 February 2025 with the Low Resolution Imaging Spectrometer \citep[LRIS,][]{Oke_LRIS_Red_1995PASP, McCarthy_LRIS_blue_1998SPIE, Rockosi_LRIS_Red_Upgrade_2010SPIE}. The spectrum spans 0.4 $\mu$m to $1.0 \ \mu$m at $R = 400$ and has SNR increasing with wavelength from 20 in the blue to 200 in the red. The PA of the slit was set to $-45^\circ$ to avoid the optical nebula to the southeast. The spectrum was extracted and flux-calibrated using the LRIS pipeline in \texttt{PypeIt} \citep{prochaska_pypeit_2020JOSS}. 

We scale the final spectrum by a factor of 4.0 to match the flux level of the IRTF/SpeX spectrum at 1.0 $\mu$m. 

\subsubsection{NIR (Palomar/TripleSpec and IRTF/SpeX)}

We use the 30 October 2015 NASA Infrared Telescope Facility (IRTF) SpeX SXD/LXD spectrum published in \citet{connelley_near-infrared_2018}. The spectrum spans $0.7-4.2 \ \mu$m at high signal-to-noise and is shown in Figure \ref{fig:FullModel}.
%\lah{[the rest of this paragraph can be eliminated since you did not follow these steps.]}

%The spectrum was extracted, flux-calibrated, and telluric-corrected using \texttt{spextool}. A final flux-calibration was performed by scaling the spectrum to simultaneous MKO $K$-band photometry taken with the SpeX guider camera. A detailed description of the procedure is given in \citet{connelley_near-infrared_2018}. 

%We obtained a NIR spectrum on 19 January 2025 using the TripleSpec instrument \citep{Herter_TripleSpecInstrument_2008SPIE} on the Palomar Observatory 200 inch Hale Telescope. The spectrum spans $0.95-2.5 \ \mu$m with an SNR of 30 at 1 $\mu$m and 400 at 2.2 $\mu$m. The spectra were taken in $3\times$ABBA nod sequences with 30 s individual exposures. We extracted, coadded, and telluric-corrected the spectra using $\mathtt{spextool}$ \citep{Cushing_spextool_2004PASP}. 

%We account for slit losses during the observation by scaling the spectrum to the near-simultaneous $J$ band flux for the object reported by the Palomar/Gattini-IR time domain survey \citep{moore_gattini_instrument_2016SPIE, Moore_Gattini_1_2019NatAs, De_Gattini_results_2020PASP}.

\subsection{Photometry}

We obtained multiband photometry of V883 Ori from several archival datasets. The target appears reliably in the Two-Micron All Sky Survey \citep[2MASS,][]{Cutri_2MASS_2003yCat}, Wide-field Infrared Survey Explorer \citep[WISE/NEOWISE,][]{WISE_Mission_2010AJ,Mainzer_neowise_2011ApJ, cutri_neowise_supplement_2015nwis}, and Zwicky Transient Facility \citep[ZTF,][]{Bellm_ZTFReference_2019PASP} point source catalogs. We also perform new aperture photometry on the source using the Pan-STARRS $grizy$ stacked images \citep{chambers2019panstarrs1surveys, Magnier_PS_dataProcessing_2020ApJS, Waters_PSStacking_2020ApJS, Flewelling_PS_database_2020ApJS} to obtain consistent and reliable measurements of the optical flux for comparison with our Keck/LRIS spectrum. The aperture photometry procedure is described in Appendix \ref{app:AperturePhot}.

To contextualize the duration of the outburst and the current state of V883 Ori relative to past observations, we also assemble 20 years of ground-based photometry in $V$, $R$, and $I$ bands. The data reduction and aperture photometry procedure for these are described in Appendix \ref{app:HistoricPhot} and the lightcurve we produce is discussed in Section \ref{sec:LC}. While the source flux is reported in the ZTF stacked image catalog, the individual detections are unreliable, so we preform aperture photometry on individual ZTF image frames to include in our lightcurve. The ZTF photometry is also described in Appendix \ref{app:HistoricPhot}.

The NEOWISE data are binned by visit and each point in the lightcurve is the 3 sigma-clipped median flux for the observations in the visit, with the uncertainty given by the 3 sigma-clipped standard deviation of the measurements. The source is extremely bright in the $W1$ and $W2$ bands, so we apply the necessary saturation corrections\footnote{\url{https://wise2.ipac.caltech.edu/docs/release/neowise/expsup/sec2_1civa.html}}. 

\section{Disk Model Fits} \label{sec:diskFits}
\begin{figure}[htb]
    \centering
    \includegraphics[trim={0 4cm 0 4cm},width=0.99\linewidth]{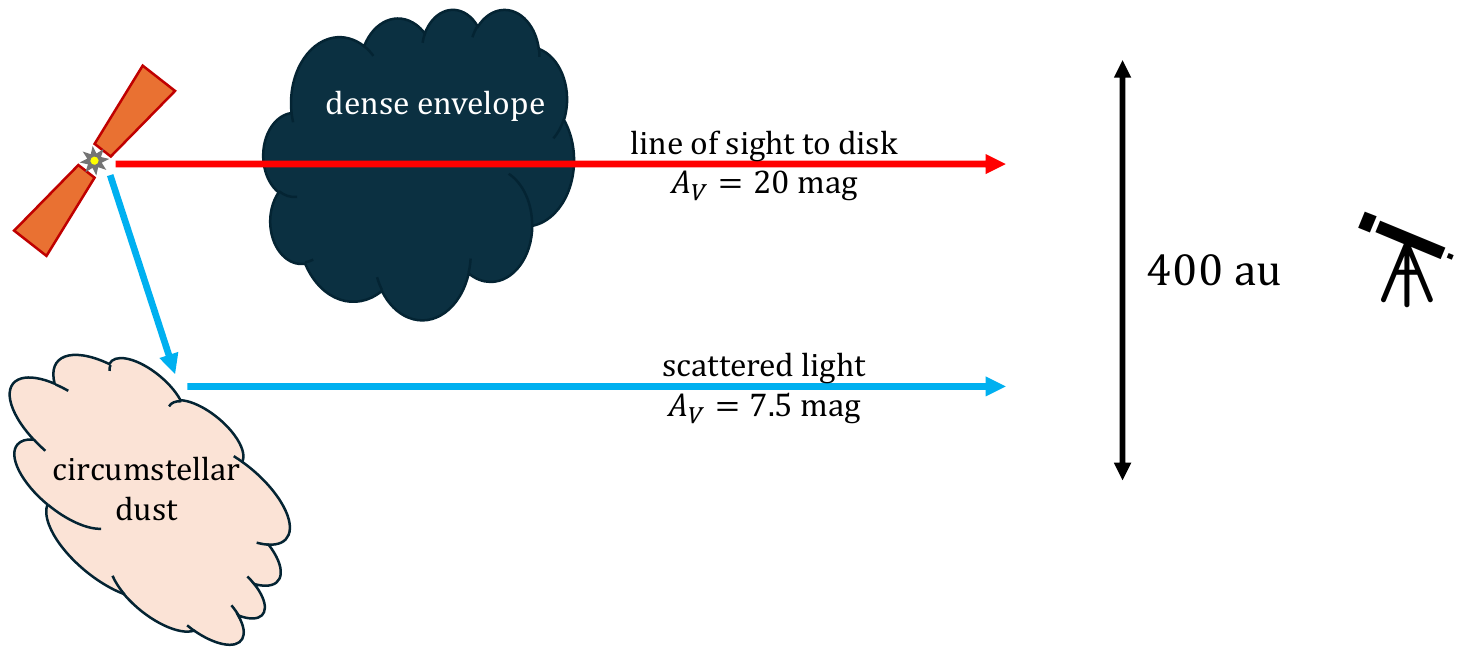}
    \caption{A diagram of the system showing the geometry of the scattering surfaces and obscuring screen relative to the disk and the observer. The illustrated disk (in orange) angle is set to the 38.3$^\circ$ disk inclination. The vertical black arrow marks the 400 au projected spatial width of the LRIS trace on-sky, while the size of the disk illustration is the 100 au dust disk radius.  }
    \label{fig:diagram}
\end{figure}

The visible/NIR spectrum of V883 Ori is shown in Figure \ref{fig:FullModel}. In this section, we describe an accretion disk model fit and demonstrate that a standard FU Ori disk model cannot fit the spectrum blueward of 1 $\mu$m. The emission in the NUV/visible is consistent with a less extincted disk, indicating it is dominated by scattered light with less line-of-sight extinction, while the NIR likely probes a line of sight directly to the inner disk. A cartoon illustrating the scenario is shown in Figure \ref{fig:diagram}.

%\lah{[not sure if this is the correct place, vs elsewhere, but could mention the 3.1um ice absorption, which is a feature of high extinction -- see discussion in CR2018.]}
%\asc{[I've added a sentence to mention this after the best-fit results.]}
We fit the entire $0.4-2.5$ $\mu$m spectrum simultaneously with a thin, viscously heated accretion disk and a physically motivated scattering law. We describe the models we adopt and fitting procedure in detail below. 

\subsection{The viscous disk}

%We fit a thin, viscously heated accretion disk model to the Palomar TripleSpec spectrum of V883 Ori from 0.9 $\mu$m to 1.8 $\mu$m. As we discuss in Section \ref{sec:scattered}, the flux observed blueward of 0.9 $\mu$m is mostly due to scattered light from the outer disk. The continuum in $K$ band, redward of 1.8 $\mu$m contains significant contributions from the passive disk, as is seen in other FUOrs like V960 Mon and HBC 722 \citep{carvalho_V960MonSpectra_2023ApJ, Carvalho_HBC722_2024ApJ}. 

The disk model we fit is a thin, viscously heated accretion disk following the formulation of \citet{Shakura_sunyaev_alpha_1973A&A} modified to be isothermal interior to $r = \frac{49}{36} R_\mathrm{inner}$ \citep{Kenyon_FUOri_disks_1988ApJ}. The details of the model are given in \citet{carvalho_V960MonSpectra_2023ApJ} and \citet{Carvalho_HBC722_2024ApJ}.  

For the distance to the source, we assume V883 Ori is a member of the Orion Nebula Cluster \citep[ONC,][]{Roychowdhury_FUOri_v883Ori_dists_2024RNAAS} and adopt the Gaia-derived distance to the ONC of 388 pc \citep{Kounkel_LamOriDist_2018AJ}. The parameters we vary in this part of the model are $\dot{M}$, $M_*$, $R_\mathrm{inner}$, the inclination ($i$), and the line of site extinction \citep[$A_V$, using][]{fitzpatrick_extinction_1999PASP}. V883 Ori has an extinction-corrected $J - K = 1.25$ magnitudes, so following the guidance of \citep{carvalho2024b} we adopt $R_\mathrm{outer} = 300 \ R_\odot$. Above this value, the $R_\mathrm{outer}$ parameter does not affect the NIR continuum in the model \citep[see Figure 6 in][]{Carvalho_V960MonPhotometry_2023ApJ}. The choice of $R_\mathrm{outer}$ is also justified by the good match to the $3-4.2 \ \mu$m region of the SpeX LXD spectrum (see Figure \ref{fig:FullModel}). 

Attempting to fit an accretion-disk-only model to the entire SED is impossible due to the drastically different slopes of the visible and NIR. As can be seen in the bottom panel of Figure \ref{fig:FullModel} and in Figure 3 of \citet{connelley_near-infrared_2018}, de-reddening the spectrum to the best-fit extinction value based on the NIR ($\sim 20$ mag) results in an extremely steep de-reddened visible range spectrum that is $10^6$ times brighter at 0.4$\mu$m than at 1 $\mu$m and gives a nonphysical $L = 8\times 10^7 \ L_\odot$. Fitting an accretion disk model to only the $1.0-2.5 \ \mu$m region of the spectrum is possible and yields best-fit parameters that are consistent with those we report for our fiducial model but are not as well-constrained.

\subsection{Scattered Light Model} \label{sec:scattered}
%Due to the enormous central source luminosity \lah{[thus far there has not been mention of a luminosity other than the discredited one just above.  aside from this, maybe instead of appealing to "enormous L" to justify scattering, try a basic physics argument here, like that there is always absorbed and scattered light, and in certain high-Av environments and with the right geometry, the scattering can be quite prominent??]} 
For a source with the right inclination relative to the observer or surrounding envelope material and significant line-of-sight extinction to the inner disk, the scattered light from the disk becomes non-negligible. In the case of the large $A_V \sim 20$ mag estimated from NIR observations of V883 Ori \citep{StromStrom_V883OriDiscovery_1993ApJ, connelley_near-infrared_2018}, the direct line-of-sight visible range flux is suppressed and must be dominated by scattering. Modeling the scattering is necessary to match the observed SED, which cannot be fit by a disk model alone (see Figure \ref{fig:FullModel}).

We use a simple 2 component scattered light model, assuming that the two physical phenomena dominating the scattering at these wavelengths is Rayleigh scattering and Mie scattering. We model the scattering of emission from the inner disk, $F_\mathrm{disk}(\lambda)$ into our line of sight as a sum of the two scattering laws,
\begin{equation} \label{eq:scatt}
    F_\mathrm{scatt}(\lambda) = F_\mathrm{disk}(\lambda) \times \left[ f_\mathrm{R} \left(\frac{9000 \mathrm{\AA}}{\lambda} \right)^{4} + f_\mathrm{Mie} \right].
\end{equation}
The fraction of Rayleigh scattering is given by $f_R$, while the fraction of light subject to Mie scattering is given by $f_\mathrm{Mie}$. The normalization of the Rayleigh component at 9000 \AA\ is chosen as the wavelength where the scattering component begins to dominate, but is formally degenerate with the value of $f_R$. The two factors $f_R$ and $f_\mathrm{Mie}$ treat only the total emergent scattered light flux along the line of sight of the observer. The model does not consider the effect of scattering phase function, since the scattered light is unresolved in our observations. For the sub-micron sized grains dominating the scattering, the Mie scattering opacity between $\lambda=0.3\mu$m and $\lambda=1.0\mu$m of most dust compositions is approximately constant \citep{Ysard_OpticalPropertiesOfDust_2018A&A}. After computing the scattered light spectrum, we redden it to $A_V(\mathrm{scatt})$, which is distinct from the $A_V$ to the viscous accretion disk. 

%\lahcomm{should figure 9 come much earlier, so as to make all of this much more clear?  i could go in 3.0 for example.}

\subsubsection{Fitting process and results}

The fit was performed using a log-likelihood minimizing Markov-Chain Monte Carlo (MCMC) technique with the nested sampling code $\mathtt{dynesty}$ \citep{speagle2020}. We adopt uniform priors for 6 of the 8 total model parameters. The two priors we impose on the disk model MCMC fits are that we require that $i$ be drawn from a normal distribution with mean $\mu = 38.3^\circ$ and standard deviation $\sigma = 3^\circ$ and that the maximum Keplerian broadening in the disk, $v_\mathrm{max} = \sqrt{GM_*/R_\mathrm{inner}} \sin i$ be drawn from a normal distribution with $\mu = 130$ km s$^{-1}$ and $\sigma = 5$ km s$^{-1}$. The prior on $i$ comes from the inclination of the outer disk, measured from exquisite millimeter imaging of the source \citep{Cieza_ALMA_V883Ori_2016Natur}. The $v_\mathrm{max}$ constraint is based on our measurements of $v_\mathrm{max}$ in the disk using the NIRSPEC spectrum and we describe the procedure in more detail in Carvalho (2025, in prep).%Appendix \ref{app:HBandCCFs}.

\begin{figure}[htb]
    \centering
    \includegraphics[width=0.98\linewidth]{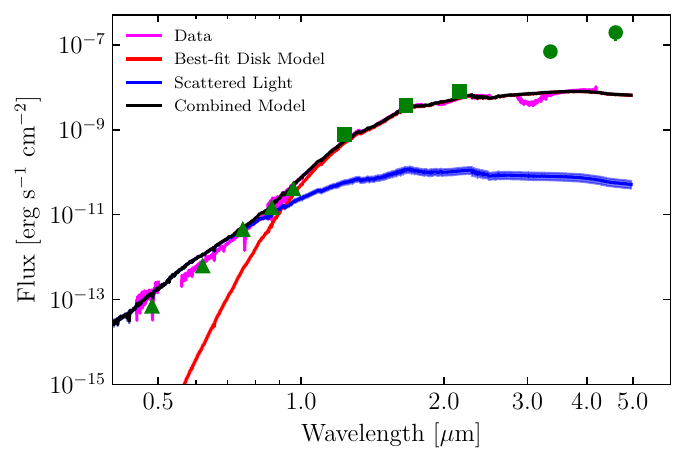}
    \includegraphics[width=0.98\linewidth]{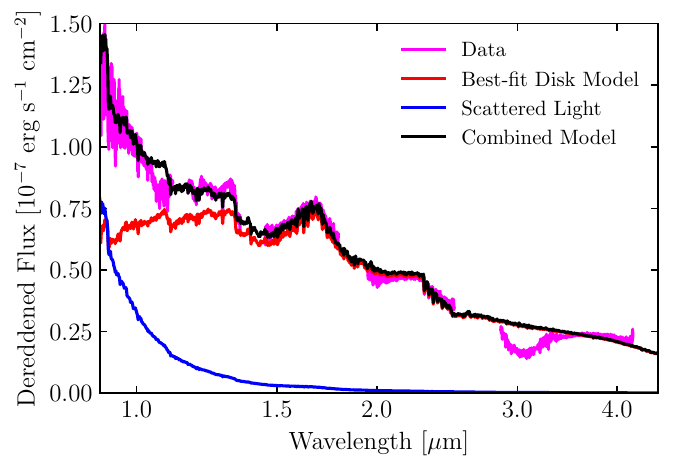}
    \caption{\textbf{Upper panel:} The complete disk $+$ scattering model (black) compared with the LRIS and SpeX spectra (magenta). The disk component is shown in red and the scattered light component shown in blue. Historical photometry of the source is shown as green symbols: triangles for PanSTARRS, squares for 2MASS, and circles for WISE. The WISE photometry includes significant thermal emission from the dust surrounding V883 Ori, while the SpeX LXD spectrum is dominated by emission from only the inner disk. \textbf{Lower panel:} The dereddened SpeX spectrum and accompanying dereddened best-fit models, to better highlight the NIR fit. The $A_V$ used for dereddening the spectrum is 20.78 mag. As established from the joint model, this is much greater than $A_V$ (scatt), causing the rapid upward slope toward the visible.}
    \label{fig:FullModel}
\end{figure}

Despite the large number of parameters, they are all extremely well-constrained and the fit to the $0.4-2.5$ $\mu$m spectrum is excellent, as can be seen in Figure \ref{fig:FullModel}. The best-fit parameters of the model are summarized in Table \ref{tab:best_fits}. The posterior distribution for each parameter is shown in a \texttt{corner} plot in Appendix \ref{app:Corner}. Our best-fit parameters for the inner disk are $\dot{M} = 10^{-3.89} \ M_\odot$ yr$^{-1}$, $M_* = 1.34 \ M_\odot$, $R_\mathrm{inner} = 5.86 \ R_\odot$, which produce $L_\mathrm{acc} = 458 \ L_\odot$ and $T_\mathrm{max} = 7046 \ L_\odot$. The best-fit inclination is $i = 38.15^\circ$ and the disk is seen through $A_V = 20.78$ mag. The large extinction to the inner disk is confirmed by the strong 3 $\mu$m ice absorption \citep{connelley_near-infrared_2018}, which can be seen in Figure \ref{fig:FullModel}. The scattered light component is seen through $A_V = 7.84$ mag, indicating it follows a less extincted line-of-sight than the inner disk. 

\begin{figure}[htb]
    \centering
    \includegraphics[width=0.9\linewidth]{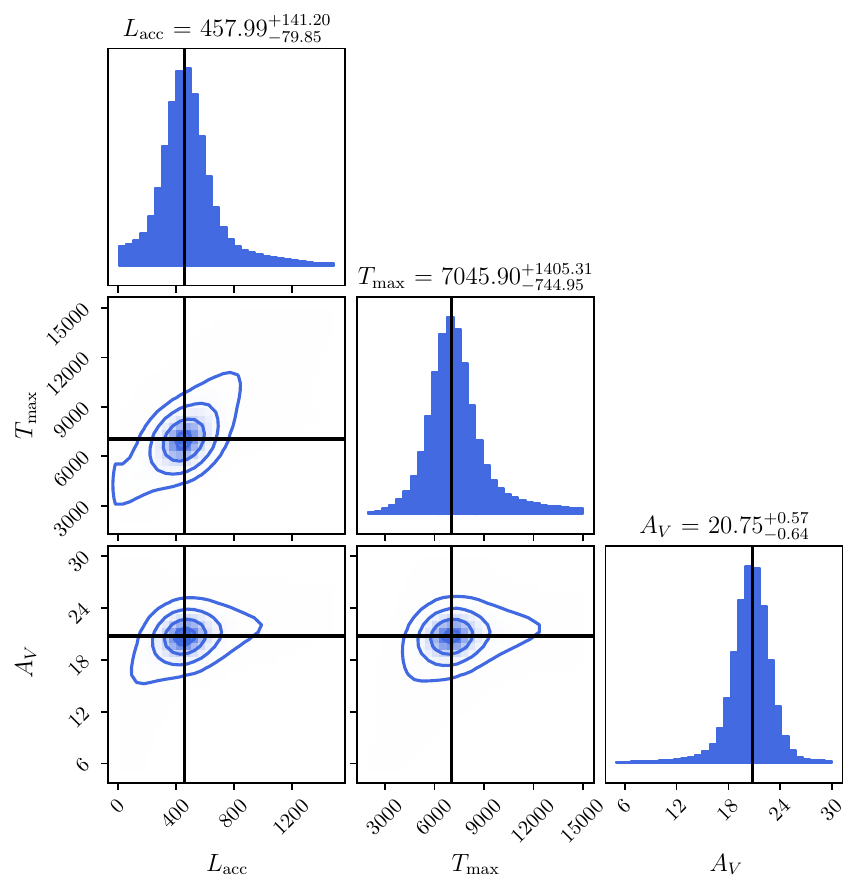}
    \caption{Posterior distributions for $L_\mathrm{acc}$ and $T_\mathrm{max}$ computed from the distributions of $M_*$, $\dot{M}$, and $R_\mathrm{inner}$. The posterior distribution for $A_V$ for the disk component is shown for reference, as it is often covariant with $T_\mathrm{max}$. The lack of strong covariance our posteriors indicates we have successfully broken the degeneracy typically present in SED fitting of accretion disk models.}
    \label{fig:LaccTmax}
\end{figure}

We also used the the posterior distributions of the disk model parameters to construct the distributions of $L_\mathrm{acc}$ and $T_\mathrm{max}$ values of the system. The distributions are shown in Figure \ref{fig:LaccTmax}. The median values of $L_\mathrm{acc} = 458 \ L_\odot$ and $T_\mathrm{max} = 7045$ K are reported in Table \ref{tab:best_fits} along with the upper and lower uncertainties.

\begin{deluxetable}{c|c|c|c}%[!htb]
	\tablecaption{Best-fit values for the \texttt{dynesty} MCMC disk $+$ scattering model.
 \label{tab:best_fits}}
	\tablewidth{0pt}
	\tablehead{
	   % \colhead{Filter} & \colhead{$\lambda_\mathrm{ref}$} & \colhead{$c_0$} & \colhead{$c_1$} & \colhead{$c_2$} & \colhead{$A_{\lambda, disk} / A_V$}
	    \colhead{Parameter} & \colhead{Symbol} & \colhead{Value} & \colhead{Units} 
     %\\
    	%    \colhead{} & \colhead{} & \colhead{mag} & \colhead{} & \colhead{mag$^{-1}$} & \colhead{}
	}
\startdata
\hline
\multicolumn{4}{c}{Disk Model} \\
\hline
Accretion Rate & log$(\dot{M})$ & $-3.89^{+0.22}_{-0.14}$ & dex $M_\odot$ yr$^{-1}$ \\
Stellar Mass & $M_*$ & $1.34^{+0.37}_{-0.46}$ & $M_\odot$ \\
Disk Inner Radius  & $R_\mathrm{inner}$ & $5.86^{+0.94}_{-1.83}$ & $R_\odot$ \\
Inclination & $i$ & $38.15^{+3.02}_{-2.93}$ & deg \\
Extinction & $A_V$ & $20.78^{+0.77}_{-0.67}$ & mag \\
\hline
\multicolumn{4}{c}{Scattered Light Model} \\
\hline
Rayleigh Factor & log$f_R$ & $-3.32^{+0.44}_{-0.45}$ &  $\cdots$ \\
Mie Factor & log$f_\mathrm{Mie}$ & $-2.31^{+0.07}_{-0.20}$ & $\cdots$ \\
Extinction & $A_V$(scatt) & $7.84^{+0.96}_{-0.57}$ & mag \\
\hline
\multicolumn{4}{c}{Inner Disk Properties} \\
\hline
Accretion Luminosity & $L_\mathrm{acc}$ & $458^{+141}_{-80}$ &  $L_\odot$ \\
Max Disk Temperature & $T_\mathrm{max}$ & $7046^{+1405}_{-744}$ &  K \\
\enddata
%    \tablenotetext{*}{We compute the absorption-only optical depth using $\tau_\nu = -\mathrm{ln}\left( 1 - A_\mathrm{dec}/B_\nu(\mathrm{11 \ K})  \right)$}
\end{deluxetable}

\section{Discussion}

Using Keck/LRIS and IRTF/SpeX, we are able to observe the inner disk of V883 Ori and constrain its accretion luminosity and maximum disk temperature, along with the physical parameters of the system. The non-negligible contribution from scattering in the visible range required modeling both the inner disk and the scattered light from the inner disk simultaneously. The combined model described in Table \ref{tab:best_fits} is an excellent match to the spectrum of the system from $0.4 - 4.2 \ \mu$m.  

The best fit stellar mass and disk inclination are both in good agreement with the $M_* = 1.3 \ M_\odot$ and $i = 38.3^\circ$ derived from millimeter range observations of the system \citep{Cieza_ALMA_V883Ori_2016Natur}. The $R_\mathrm{inner}=5.86 \ R_\odot$ for the system is much larger than the $2.7-3.5 \ R_\odot$ radii expected for a 0.5-1 Myr old system \citep{Baraffe_isochrones_2015A&A}, but FUOrs are likely to have inflated radii due to their rapid mass accretion \citep{hartmann_fu_1996}. The ability of rapid mass accretion to inflate the radius of a star is seen in models of main sequence accretors in symbiotic binary systems \citep{2024ApJ...966L...7L}.

\subsection{The scattered light component} \label{sec:scatt_disc}

The visible range spectroscopy and spectrophotometry of V883 Ori are both many orders of magnitude brighter than is expected for a FUOr inner disk observed through $A_V = 20$ mag. The disk model shown in Figure \ref{fig:FullModel} indicates that the source should barely be detectable blueward of 0.6 $\mu$m. Nevertheless, we see faint emission from V883 Ori across the whole visible range in the Pan-STARRS photometry and the Keck/LRIS spectrum, indicating a different emission component in the system must be dominating the visible spectrum. 

The optical spectrum is also consistent with that of a FUOr, as was first reported in \citet{Allen_V883Ori_1975MNRAS}. We affirm this with our LRIS spectrum. It is therefore more likely that the blue excess is due to the dust surrounding V883 Ori scattering the optical spectrum of the inner disk into our line of sight. This requires scattering only 0.5\% of the radiation, which is typical of the fraction of starlight scattered by typical circumstellar disks \citep[e.g.,][]{Wolff_scatteredLightDisk_2017ApJ}. 

The source has been observed at high ($\sim 10 \ \mu$Jy) sensitivity with ALMA at angular resolutions spanning $0.04''-1.0''$ and fields of view from $0.3'-1.0'$, covering both the point source and the extended nebula \citep{cieza_v883Ori_2018MNRAS, Lee_V883Ori_COMs_2019NatAs, Tobin_v883Ori_2023Natur}. Matching the visible range excess emission with a stellar companion would require a luminosity of $L > 2.25 \ L_\odot$ assuming $A_V \sim 7.8$ and a temperature of $\sim 10,000$ K. This would imply a radius of $0.5 \ R_\odot$, which is unphysically small for such a hot star. Furthermore, such a companion would be expected to disrupt the V883 Ori outer disk, which would be detectable in the millimeter dust emission and no such disruption to the disk has been reported. 

Typically, in cases where the central source is highly extincted such that the visible/NIR are largely seen as scattered light, the extinction is due to the flared disk itself. These systems are edge-on, such that the disk obscures a direct line of sight to the star and the light from the star is mostly seen scattered off the side of the disk opposite the observer \citep[e.g., HH 30 and IRAS04302+2247,][]{Burrows_HH30_HST_1996ApJ,Padgett_IRAS04302_HST_1999AJ}. The outer disk inclination of V883 Ori is too face-on for this scenario. 

We propose instead that the visible light is scattered off the outer disk and nearby circumstellar material. The direct line of the sight to the inner disk is obscured by a cloud of dense envelope material, which produces the 20 magnitudes of optical extinction to the inner disk spectrum. The less extincted visible light is scattered off material near the disk but not obscured by this cloud, so we see it through a much lower $A_V$, as shown in Figure \ref{fig:diagram}.

The full-width at half maximum (FWHM) of the optical extraction window used to extract the LRIS spectrum is 1.0$''$, which translates to a physical size of $388$ au at the source location. The dust continuum emission in the disk only extends to $0.25''$ or a radius of 100 au \citep{Cieza_ALMA_V883Ori_2016Natur}. Large dust structures surrounding face-on FUOrs beyond their disks are seen in scattered light at flux ratios of $\sim1 \%$ of the inner disk luminosity \citep[e.g., V960 Mon and FU Ori,][]{Weber_V960MonSpiralsClumps_2023ApJ, Zurlo_ScatteredLight_2024A&A}. In FU Ori, the light-scattering circumstellar envelope extends more than 200 au from the source, despite the extremely compact disk radius of $\sim 11$ au \citep{Perez_FUOriALMA_2020ApJ}.

The $^{12}$CO and $^{13}$CO millimeter emission from V883 Ori reveals a biconical outflow extending 5$^{\prime\prime}$ from the disk \citep{RR_ALMA_V883Ori_outflow_2017MNRAS}. The same observations also show that within 0.5$^{\prime\prime}$ of the disk, the $^{12}$CO emission suffers significant absorption. The absorption is attributed to a large dust column along the line of sight to the inner disk, which would be consistent with the visible/NIR observations presented here. It is possible that the large outflow in the system traces a nebula near the disk that is contributing to the scattered light component, as is seen in FU Ori \citep{2024ApJ...966...96H}.

\subsection{The enormous mass consumed by V883 Ori} \label{sec:LC}
\begin{figure*}[htb]
    \centering
    \includegraphics[width=0.99\linewidth]{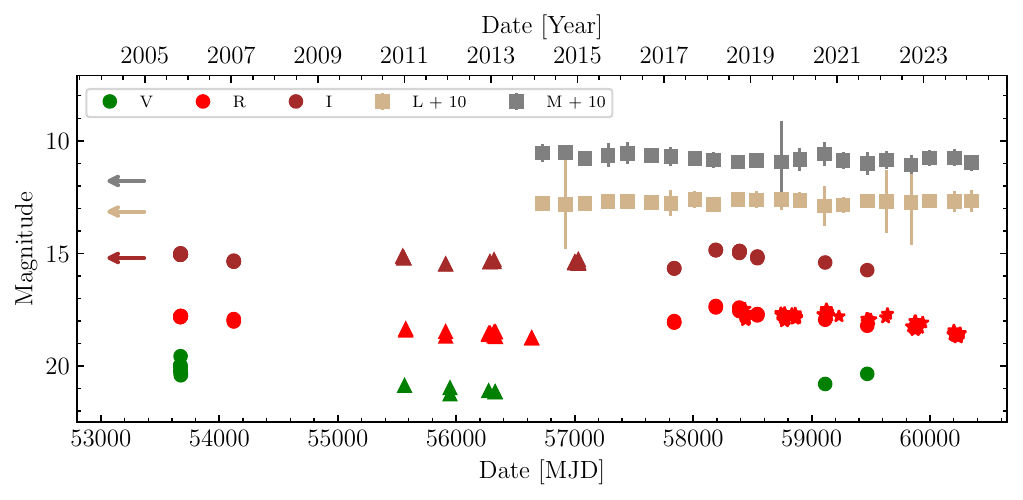}
    \caption{A lightcurve of V883 Ori, spanning the past 20 years of observations. The arrows to the left of the plot mark the brightness of the source at $0.8$, $3.5$, and $4.8$ $\mu$m, taken in 1963, 1975, and 1975, respectively \citep{Allen_V883Ori_1975MNRAS}. The source has remained at essentially the same brightness for at least the past 60 years. The triangles mark points from Pan-STARRS, while $*$ symbols mark points from ZTF. The tan and grey squares mark NEOWISE $W1$ and $W2$ points, respectively. The symbols on the plot are larger than the photometric uncertainties for all points except NEOWISE.}
    \label{fig:V883OriLC}
\end{figure*}

The luminosity of V883 Ori has remained essentially constant for at least the past 60 years of observation, and plausibly longer, implied by Figure \ref{fig:V883OriLC}. This indicates that the mass accretion rate has remained at an extremely high $10^{-3.9} \ M_\odot$ yr$^{-1}$ that entire time. Over 60 years, the source has accreted a minimum of $8 \ M_\mathrm{Jup}$ ($\dot{M}\times \Delta t$). In their simulation of an FU Ori outburst, \citet{Zhu_outburst_FUOri_2020MNRAS} found that the mechanical energy removal by winds during the burst can be highly efficient, resulting in a 30\% lower $L_\mathrm{acc}$ than the thin disk model predicts. Thus, for a given $L_\mathrm{acc}$, the actual $\dot{M}$ in the disk can be $3\times$ greater. If this is the case in V883 Ori, the total accreted mass in 60 years could be as much as 24 $M_\mathrm{Jup}$. 

The origin of the outburst is estimated to be around 1888, when the nebula was first reported by \citep{Pickering_V883OriNebula_1890AnHar}. \citet{StromStrom_V883OriDiscovery_1993ApJ} argue that given the faint $B \sim 24$ mag arcsec$^{-2}$ surface brightness of the nebula in 1975 and the fact that plates in 1888 were preferentially sensitive to blue light and much less sensitive generally, the nebula must have been much brighter. If the star is illuminating the nebula sufficiently that the nebula would be brightened by the outburst, then its accretion rate would have been even greater in the 1880's. If we conservatively assume that the accretion rate was at least the same at outburst as it is today, then the star has accreted at least 18 $M_\mathrm{Jup}$ (0.017 $M_\odot$). This is comparable to lower estimates for the total current disk mass \citep[e.g., $0.02 \ M_\odot$,][]{Sheehan_VANDAM_V883Ori_2022ApJ}, although the disk mass has also been estimated to be as high as 0.54 $M_\odot$ \citep[in radiative transfer modeling of dust continuum emission and adopting a dust-to-gas ratio of 0.01,][]{cieza_v883Ori_2018MNRAS}. 

Even if the disk mass is underestimated, the total mass accreted by V883 Ori during this outburst may be a large fraction of the mass in the disk. FU Ori is a similar case. The system has been in outburst since 1937 \citep{Wachmann_FUOri_1954ZA} and has only faded by $\sim 1$ mag since the burst. Today, the accretion rate is reported to be $10^{-4.49}$ $M\odot$ yr$^{-1}$ \citep{Carvalho_FUVSpectrumFUOri_2024ApJ}. Assuming that the accretion rate has been constant in the interesting of obtaining a conservative estimate of mass accreted, the total is 0.003 $M_\odot$, of almost half of the total current $M_\mathrm{disk} = 0.007 M_\odot$ \citep[using the dust disk mass from][and assuming a dust-to-gas ratio of 0.01]{Perez_FUOriALMA_2020ApJ}.

\subsection{Irradiation heating of the outer disk} \label{sec:irradiation}
As one of the closest FUOrs to the Sun, V883 Ori has been the target of several millimeter-range observations to study the molecular emission from the spatially-resolved outer disk. Interpreting the observed spectra, however, is challenging because it requires careful accounting for all potential heating sources in the disk and the disk environment. 

One critical heating source in FUOr disks is the irradiation from the self-luminous inner disk. The extremely high accretion luminosity in the inner disk dominates the UV to NIR flux.  As we demonstrate below, this accretion luminosity also dominates the irradiation of the outer disk, whereas the central star's luminosity is irrelevant, and its irradiative effects inconsequential in the outer disk. 

To investigate the impact of omitting inner disk irradiation from a heating budget of the outer disk, we construct an analytic heating model based on that of \citet{Fukue_Irradiation_2013PTEP}. This model is distinct from that which we used to fit the visible/NIR SED of the source in Section \ref{sec:diskFits} and is meant to illustrate the different contributions of accretion and irradiation to regions of the disk probed by millimeter-range observations. 
We compute the irradiation heating from both the star and inner disk as
\begin{equation} \label{eq:Qirr_star}
    Q_\mathrm{irr,*} = (1 - A) \left[ \frac{R_*L_*}{3\pi^2 r^3} + \frac{L_*}{4\pi^2 r^2}\left( \frac{dH}{dr} - \frac{H}{r} \right) \right],
\end{equation}
and
\begin{equation} \label{eq:Qirr_disk}
    Q_\mathrm{irr,disk} = (1 - A) \left[\frac{L_\mathrm{acc}}{4\pi^2 r^2}\left( \frac{dH}{dr} - \frac{H}{r} \right) \right].
\end{equation}
In Equation \ref{eq:Qirr_star}, the first term $\frac{R_*L_*}{3\pi^2 r^3}$ accounts for the physical extent of the star, which slightly impacts the heating at $r < 10$ au. At larger radii, the star can be treated like a point source and the radiation angle of incidence on the disk, accounted for by the second term $\frac{L_*}{4\pi^2 r^2}\left( \frac{dH}{dr} - \frac{H}{r} \right)$, dominates. For the irradiation from the inner disk in Equation \ref{eq:Qirr_disk}, we drop the first term since the disk is flat and the angle of incidence to small radii becomes extremely small. We assume that the albedo of the outer disk, $A$, is 0 for simplicity\footnote{Our scattering model indicates it is not 0, but given the very small scattering factors, it is close enough to not meaningfully affect the calculation here.}. We adopt a stellar radius of $R_* = 2.7 \ R_\odot$ and $L_* = 2.95 \ L_\odot$, which correspond to a $M_* = 1.3 \ M_\odot$ star at an age of 1 Myr \citep{Baraffe_isochrones_2015A&A}. We assume that the disk scale height, $H(r) = H_0 \left( r/r_0 \right)^{\gamma}$, where $H_0 = 0.05$ au at $r_0 = 1$ au and the flare index $\gamma = 1.1$. For the disk $L_\mathrm{acc}$ we use our best-fit value of $458 \ L_\odot$. 

\begin{figure}[!htb]
    \centering
    \includegraphics[width=0.9\linewidth]{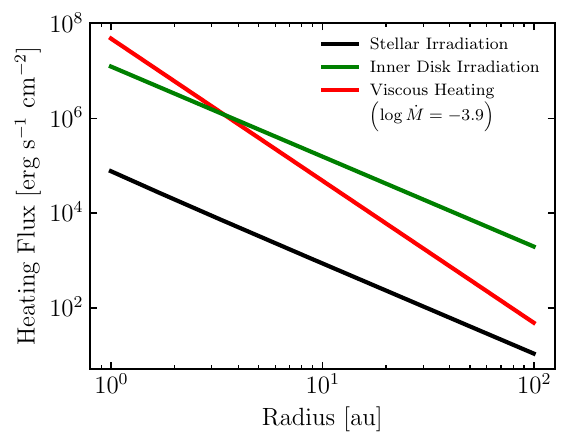}
    \caption{The sources of heating in the V883 Ori disk. The heating from the central star is negligible since the disk dominates the luminosity of the system. The midplane viscous heating, while dominant in the inner 3 au of the disk, falls below the heating due to irradiation from the accretion luminosity of the disk. }
    \label{fig:heating}
\end{figure}

We also compare this to the viscous heating due to the high accretion rate in the system, which we model following the \citet{Shakura_sunyaev_alpha_1973A&A} model,
\begin{equation}
    Q_\mathrm{visc} = \frac{3 G M_* \dot{M}}{8 \pi r^3}.
\end{equation}
The three heating models are shown in Figure \ref{fig:heating}. While the viscous heating is dominant in the inner 2 au of the disk, by $r = 4$ au the irradiation from the inner disk is greater and by $r = 25$ au, $Q_\mathrm{irr,disk} = 10 \ Q_\mathrm{visc}$. 

We note that the calculations above treat the inner disk as a luminous point source. A more detailed analysis that incorporates the radially extended nature of the inner disk and the true angle of incidence of the accretion flux on the flared outer disk is necessary to account for the correct outer disk dust temperature.

\section{Conclusion}

We use optical $+$ NIR high and low resolution spectra to constrain a viscous accretion disk model for one of the brightest and longest-lived FU Ori outbursts, V883 Ori. We successfully model the spectrum from $0.4 - 4.2$ $\mu$m by assuming that the shorter wavelength component is due to scattered light from the inner disk observed at a lower line-of-sight extinction, $A_V(\mathrm{scatt}) = 7.84$ mag, relative to the $A_V = 20.78$ mag along direct line-of-sight to the inner disk. The best-fit physical parameters of the disk model are $M_* = 1.34 \ M_\odot$, $\dot{M} = 10^{-3.89} \ M_\odot$ yr$^{-1}$, $R_\mathrm{inner} = 5.86 \ R_\odot$, and $i = 38.15^\circ$. The maximum temperature of the inner disk, given these parameters, is 7045 K and the accretion luminosity is 458 $L_\odot$. 

We also provide a decades-long light curve of the source, demonstrating its consistent brightness in both the visible and NIR over the past 62 years of observations. The flat lightcurve indicates a relatively constant accretion rate, which can be integrated to show that the source has accreted at least $8 \ M_\mathrm{Jup}$ since 1963 and has likely ejected $\sim 1 \ M_\mathrm{Jup}$ of material as outflows. 

The inner disk maximum temperature and luminosity of the source should be considered when modeling both dust continuum emission and molecular emission. The tight constraints we provide for both will enable accurate modeling of the dust and gas temperature, and therefore more reliable estimates of the disk mass of V883 Ori. Given the large mass already accreted by the source, and competing theories for trigger mechanisms of FU Ori outbursts, an accurate measurement of the disk mass is critical to understanding the past and predicting the future evolution of this object. 

\section{Acknowledgements}

The operation of the RC80 and Schmidt telescopes at Konkoly Observatory has been supported by the GINOP 2.3.2-15-2016-00033 and GINOP-2.3.2-15-2016-00003 grants of the National Research, Development and Innovation Office (NKFIH) funded by the European Union.

This work was also supported by the NKFIH NKKP grant ADVANCED 149943. Project no.149943 has been implemented with the support provided by the Ministry of Culture and Innovation of Hungary from the National Research, Development and Innovation Fund, financed under the NKKP ADVANCED funding scheme. 

\bibliography{references}{}
\bibliographystyle{aasjournal}

\appendix 

\restartappendixnumbering

\section{Aperture photometry on the Pan-STARRS stacked images} \label{app:AperturePhot}
Due to the nebulosity in the vicinity of V883 Ori, the PSF photometry of the source in the Pan-STARRS point source catalog is unreliable for bands $izy$. In order to complete the photometric visible/NIR SED of V883 Ori, we downloaded $grizy$ $2'\times 2'$ cutouts from the image cutout service \footnote{\url{https://ps1images.stsci.edu/cgi-bin/ps1cutouts}}. We then used the \texttt{photutils} package \citep{larry_bradley_2024_13989456} to extract the flux from the location of V883 Ori using a circular aperture with a $3''$ diameter. We estimate the background as the median pixel value in the image and subtract the total background counts within the aperture. We convert the total data number ($DN$) in our aperture to a magnitude using the formula in the Pan-STARRS stack images documentation\footnote{\url{https://outerspace.stsci.edu/display/PANSTARRS/PS1+Stack+images}}: $m_\lambda = -2.5 \log_{10}(DN) + 25 + 2.5\log_{10}(t_\mathrm{exp})$. The new photometry we find for V883 Ori (in AB magnitudes) is: $m_g = 21.27$, $m_r = 18.64$, $m_i = 16.27$, $m_z = 14.84$, $m_y = 13.62$. The conversions to Vega magnitudes, computed from the ratio of Vega zeropoints to AB zeropoints, are $-0.095$, 0.146, 0.37, 0.51, and 0.55, respectively. 

\section{20 years of ground-based photometry of V883 Ori} \label{app:HistoricPhot}
The visible/NIR flux level (and therefore mass accretion rate) of V883 Ori has been remarkably constant over the past 60 years. In order to constrain the photometric evolution of the source in the last 20 years, we assembled $V$, $R$, and $I$ band measurements from several ground-based facilities. The calibration, aperture extraction, and filter normalization of each data set is described in this Appendix. The aperture extraction generally follows a similar procedure as that described for the Pan-STARRS data, though our flux calibration step differs by observatory.

\subsection{ZTF}
The ZTF $r$ band individual frames were obtained from the IPAC/IRSA ZTF Image Access tool\footnote{\url{https://irsa.ipac.caltech.edu/Missions/ztf.html}}.
We selected a 3 pixel aperture radius, which requires the smallest correction between aperture and PSF fitting photometry, according to the ZTF Explanatory Supplement. We again estimate the background as the median flux level in the image. We then extract the flux at the location of V883 Ori, apply the magnitude zero point of the image specified by the \texttt{MAGZP} header key, and add the aperture to PSF correction specified by the \texttt{APCOR3} header key. 

To ensure the robustness of the extracted flux to aperture size choice, we also tested 4 and 5 pixel radii and found that the extracted fluxes are consistent with the 3 pixel values. 

\subsection{Konkoly Observatory, Instituto de Astrofisica de Canarias (IAC), and La Silla Observatory}

Optical observations of V883 Ori were obtained at five different telescopes in Hungary, Spain, and Chile. Three telescopes were at Konkoly Observatory (Hungary): the 1\,m RCC telescope equipped with a Princeton Instruments VersArray:1300b camera, pixel scale: $0\farcs31$/pixel, field of view: $6.9'{\times}6.4'$ ($V$, $R_C$, $I_C$ filters, data from 2005), the 90/60\,cm Schmidt telescope equipped with an Apogee Alta U16 camera, pixel scale: $1\farcs03/$pixel, field of view: $70'{\times}70'$ ($R_C$, $I_C$ filters, data from 2017--2019), and the Astrosysteme ASA800 80\,cm RC telescope equipped with an FLI Microline camera, pixel scale: $0\farcs55/$pixel, field of view: $18\farcm4{\times}18\farcm7$ ($V$, $r'$, $i'$ filters, data from 2020--2022). The telescope we used at Teide Observatory (Canary Islands, Spain) was the Instituto de Astrof\'\i{}sica de Canarias 82\,cm IAC80 telescope equipped with the CAMELOT E2V back illuminated chip, pixel scale: $0\farcs304$pixel, field of view: $10\farcm4{\times}10\farcm4$ ($R$, $I$ filters, data from 2007). In Chile, we used the 60 cm Rapid Eye Mount (REM) telescope located at La Silla Observatory, equipped with the ROS2 multi-channel imaging camera, $0\farcs58/$pixel, field of view: $9\farcm9{\times}9\farcm9$ ($g'$, $r'$, $i'$, $z'$ filters, data from 2014--2015). At each observing night, 3-4 images were taken with each filter, with exposure times between 20--1080\,s. The data was reduced following the usual steps for bias, dark, and flatfield correction.

For each image, we use an aperture with a 6 pixel radius, corresponding to an angular radius of 6$^{\prime\prime}$ on PSCH and $1-1.5^{\prime\prime}$ on RCC, RC80, and IAC80. The aperture size is chosen to minimize the chance of including emission from the nearby nebula while maximizing robustness to the highly variable seeing. We extract the flux at the location of V883 Ori and all Pan-STARRS catalog sources within 30' of V883 Ori. For PSCH, this accounts for half of the sources in the field of view, while for RCC, RC80, and IAC80, we use every Pan-STARRS source in the field. We take the background to the median flux level in the image. 

We computed the magnitude of V883 Ori in each image relative to the Pan-STARRS catalogue objects in the field, using the $g$ band values as reference for $V$ band, $r$ band as reference for $r'$, $R$, and $R_C$, and $i$ band as reference for $i'$, $I$, and $I_C$. We took the 5-sigma-clipped median and standard deviations as our magnitude measurements and uncertainties on the measurements, respectively. The typical uncertainty on a measurement is $0.05-0.1$ mag. Many of the REM $g'$ and $r'$ images are not deep enough to detect V883 Ori, so they are omitted from the lightcurve. 

To convert to a uniform set of $V$, $R$, and $I$ band measurements in the Johnson filter system, we used the relative zero points between each filter and the Pan-STARRS filters, and then between the AB and Vega systems for each filter. The zero points were all obtained from the Spanish Virtual Observatory website\footnote{\url{https://svo2.cab.inta-csic.es/svo/theory/fps3/}}. 

\begin{deluxetable}{cccccccc}
\tablecaption{The $V$, $R$, $I$, $W1$, and $W2$ band photometry of V883 Ori show in Figure \ref{fig:V883OriLC}. 
 \label{tab:photometry}}
\tablehead{\colhead{MJD} & \colhead{V} & \colhead{R} & \colhead{I} & \colhead{z} & \colhead{W1} & \colhead{W2} & \colhead{Facility}}
\startdata
53671.0419 & 19.966 & $\cdots$ & $\cdots$ & $\cdots$ & $\cdots$ & $\cdots$ & RCC \\
53671.0456 & $\cdots$ & 17.826 & $\cdots$ & $\cdots$ & $\cdots$ & $\cdots$ & RCC \\
53671.0492 & $\cdots$ & $\cdots$ & 15.055 & $\cdots$ & $\cdots$ & $\cdots$ & RCC \\
54121.9191 & $\cdots$ & $\cdots$ & 15.354 & $\cdots$ & $\cdots$ & $\cdots$ & IAC80 \\
54121.9330 & $\cdots$ & $\cdots$ & 15.364 & $\cdots$ & $\cdots$ & $\cdots$ & IAC80 \\
57839.7761 & $\cdots$ & 18.066 & $\cdots$ & $\cdots$ & $\cdots$ & $\cdots$ & PSCH \\
57839.7784 & $\cdots$ & $\cdots$ & 15.651 & $\cdots$ & $\cdots$ & $\cdots$ & PSCH \\
58429.3919 & $\cdots$ & 17.770 & $\cdots$ & $\cdots$ & $\cdots$ & $\cdots$ & ZTF \\
58432.3703 & $\cdots$ & 17.823 & $\cdots$ & $\cdots$ & $\cdots$ & $\cdots$ & ZTF \\
59114.1002 & 20.797 & $\cdots$ & $\cdots$ & $\cdots$ & $\cdots$ & $\cdots$ & RC80 \\
59114.1037 & $\cdots$ & 17.943 & $\cdots$ & $\cdots$ & $\cdots$ & $\cdots$ & RC80 \\
59114.1072 & $\cdots$ & $\cdots$ & 15.400 & $\cdots$ & $\cdots$ & $\cdots$ & RC80 \\
59265.0747 & $\cdots$ & $\cdots$ & $\cdots$ & $\cdots$ & 2.838 & 0.862 & WISE \\
59473.4018 & $\cdots$ & $\cdots$ & $\cdots$ & $\cdots$ & 2.669 & 0.996 & WISE \\
\enddata
\end{deluxetable}

\section{Posterior Distributions of the Disk $+$ Scattered Light Model} \label{app:Corner}
\begin{figure*}
    \centering
    \includegraphics[width=0.98\linewidth]{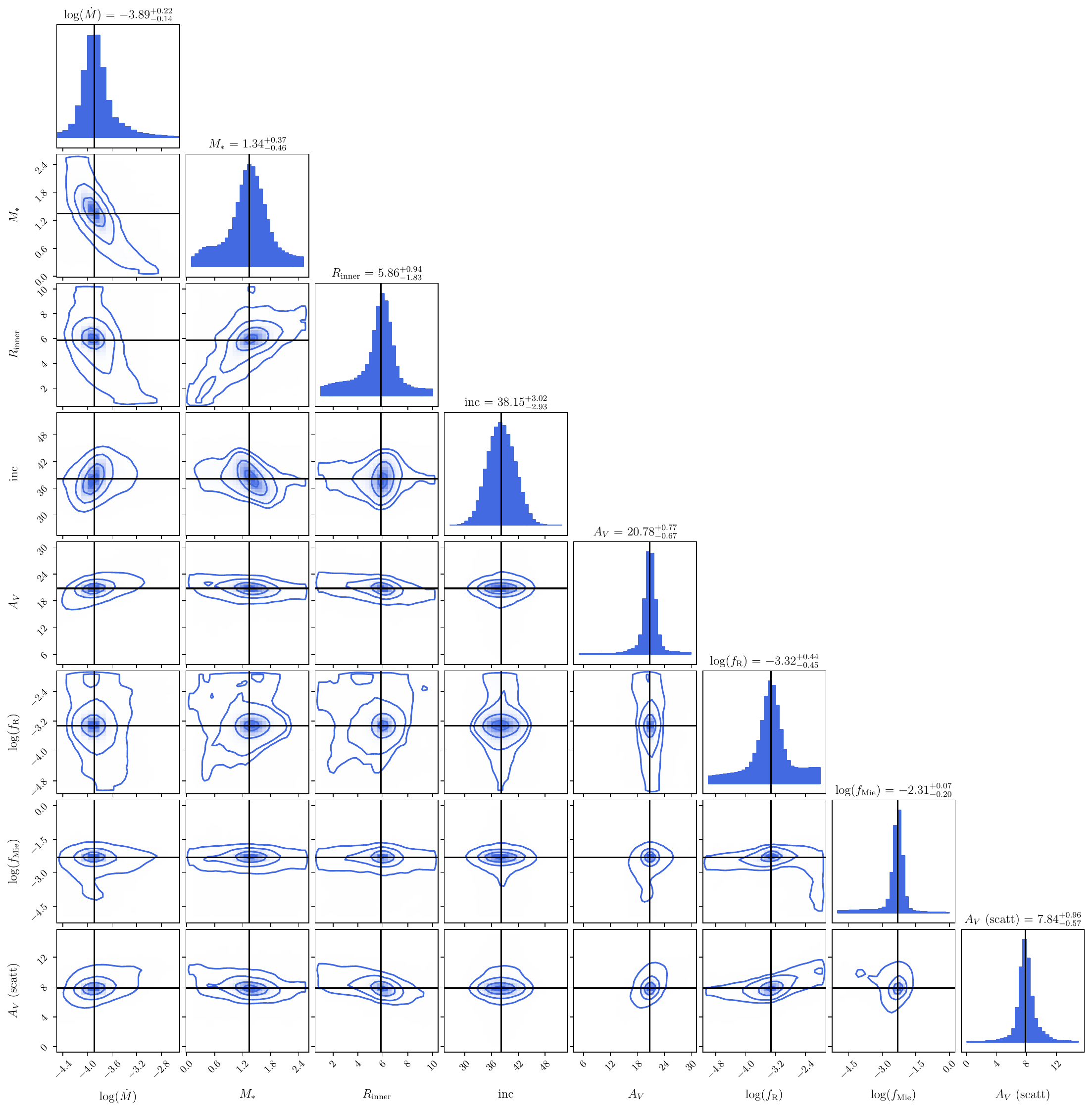}
    \caption{The posterior distributions of all 8 parameters of the disk $+$ scattering models. They are all well-constrained and show minimal covariance. } 
    \label{fig:cornerDynestyAll}
\end{figure*}

Our MCMC \texttt{dynesty} models ran for 22,306 iterations. It reached a $d$log$z$ of 0.01 and then stopping criterion of 1.0. The posterior distributions of each parameter are shown as a \texttt{corner} \citep{corner_FM_2016} plot in Figure \ref{fig:cornerDynestyAll}.

\end{document}